\documentclass[twocolumn,showpacs,preprintnumbers,amsmath,amssymb,eqsecnum]{revtex4}
\usepackage{dcolumn}
\usepackage{bm}
\usepackage[dvips]{graphicx}
\usepackage{wrapfig}
\usepackage{psfrag}
\usepackage{color}
\begin{document}


\def\sh{\mathop{\rm sh}\nolimits}
\def\ch{\mathop{\rm ch}\nolimits}
\def\var{\mathop{\rm var}}\def\exp{\mathop{\rm exp}\nolimits}
\def\Re{\mathop{\rm Re}\nolimits}
\def\Sp{\mathop{\rm Sp}\nolimits}
\def\kp{\mathop{\text{\ae}}\nolimits}
\def\bk{{\bf {k}}}
\def\bp{{\bf {p}}}
\def\bq{{\bf {q}}}
\def\lra{\mathop{\longrightarrow}}
\def\Const{\mathop{\rm Const}\nolimits}
\def\sh{\mathop{\rm sh}\nolimits}
\def\ch{\mathop{\rm ch}\nolimits}
\def\var{\mathop{\rm var}}
\def\mK{\mathop{{\mathfrak {K}}}\nolimits}
\def\mR{\mathop{{\mathfrak {R}}}\nolimits}
\def\mv{\mathop{{\mathfrak {v}}}\nolimits}
\def\mV{\mathop{{\mathfrak {V}}}\nolimits}
\def\mD{\mathop{{\mathfrak {D}}}\nolimits}
\def\mN{\mathop{{\mathfrak {N}}}\nolimits}
\def\mS{\mathop{{\mathfrak {S}}}\nolimits}
\def\mc{\mathop{{\mathfrak {c}}}\nolimits}

\newcommand\ve[1]{{\mathbf{#1}}}

\def\Re{\mbox {Re}}
\newcommand{\Z}{\mathbb{Z}}
\newcommand{\R}{\mathbb{R}}
\def\mK{\mathop{{\mathfrak {K}}}\nolimits}
\def\mk{\mathop{{\mathfrak {k}}}\nolimits}
\def\mR{\mathop{{\mathfrak {R}}}\nolimits}
\def\mv{\mathop{{\mathfrak {v}}}\nolimits}
\def\mV{\mathop{{\mathfrak {V}}}\nolimits}
\def\mD{\mathop{{\mathfrak {D}}}\nolimits}
\def\mN{\mathop{{\mathfrak {N}}}\nolimits}
\def\ml{\mathop{{\mathfrak {l}}}\nolimits}
\def\mf{\mathop{{\mathfrak {f}}}\nolimits}
\def\mt{\mathop{{\mathfrak {t}}}\nolimits}
\newcommand{\ccm}{{\cal M}}
\newcommand{\cE}{{\cal E}}
\newcommand{\cV}{{\cal V}}
\newcommand{\cI}{{\cal I}}
\newcommand{\cR}{{\cal R}}
\newcommand{\cK}{{\cal K}}
\newcommand{\cH}{{\cal H}}
\newcommand{\cW}{{\cal W}}
\newcommand{\cL}{{\cal L}}
\newcommand{\cN}{{\cal N}}

\def\br{\mathop{{\bf {r}}}\nolimits}
\def\bS{\mathop{{\bf {S}}}\nolimits}
\def\bA{\mathop{{\bf {A}}}\nolimits}
\def\bJ{\mathop{{\bf {J}}}\nolimits}
\def\bn{\mathop{{\bf {n}}}\nolimits}
\def\bg{\mathop{{\bf {g}}}\nolimits}
\def\bv{\mathop{{\bf {v}}}\nolimits}
\def\be{\mathop{{\bf {e}}}\nolimits}
\def\bp{\mathop{{\bf {p}}}\nolimits}
\def\bz{\mathop{{\bf {z}}}\nolimits}
\def\bbf{\mathop{{\bf {f}}}\nolimits}
\def\bb{\mathop{{\bf {b}}}\nolimits}
\def\ba{\mathop{{\bf {a}}}\nolimits}
\def\bx{\mathop{{\bf {x}}}\nolimits}
\def\by{\mathop{{\bf {y}}}\nolimits}
\def\br{\mathop{{\bf {r}}}\nolimits}
\def\bs{\mathop{{\bf {s}}}\nolimits}
\def\bH{\mathop{{\bf {H}}}\nolimits}
\def\bk{\mathop{{\bf {k}}}\nolimits}
\def\be{\mathop{{\bf {e}}}\nolimits}
\def\bnul{\mathop{{\bf {0}}}\nolimits}
\def\bq{{\bf {q}}}

\newcommand{\oV}{\overline{V}}
\newcommand{\vkp}{\varkappa}
\newcommand{\os}{\overline{s}}
\newcommand{\opsi}{\overline{\psi}}
\newcommand{\ov}{\overline{v}}
\newcommand{\oW}{\overline{W}}
\newcommand{\oPhi}{\overline{\Phi}}

\def\mI{\mathop{{\mathfrak {I}}}\nolimits}
\def\mA{\mathop{{\mathfrak {A}}}\nolimits}

\def\st{\mathop{\rm st}\nolimits}
\def\tr{\mathop{\rm tr}\nolimits}
\def\sign{\mathop{\rm sign}\nolimits}
\def\d{\mathop{\rm d}\nolimits}
\def\const{\mathop{\rm const}\nolimits}
\def\diag{\mathop{\rm diag}\nolimits}
\def\O{\mathop{\rm O}\nolimits}
\def\Spin{\mathop{\rm Spin}\nolimits}
\def\exp{\mathop{\rm exp}\nolimits}
\def\SU{\mathop{\rm SU}\nolimits}
\def\mU{\mathop{{\mathfrak {U}}}\nolimits}
\newcommand{\cU}{{\cal U}}
\newcommand{\cD}{{\cal D}}

\def\mI{\mathop{{\mathfrak {I}}}\nolimits}
\def\mA{\mathop{{\mathfrak {A}}}\nolimits}
\def\mU{\mathop{{\mathfrak {U}}}\nolimits}

\def\st{\mathop{\rm st}\nolimits}
\def\tr{\mathop{\rm tr}\nolimits}
\def\sign{\mathop{\rm sign}\nolimits}
\def\d{\mathop{\rm d}\nolimits}
\def\const{\mathop{\rm const}\nolimits}
\def\O{\mathop{\rm O}\nolimits}
\def\Spin{\mathop{\rm Spin}\nolimits}
\def\exp{\mathop{\rm exp}\nolimits}

\title{Unitarity of 4D Lattice Theory of Gravity}

\author {S.N. Vergeles\vspace*{4mm}\footnote{{e-mail:vergeles@itp.ac.ru}}}

\affiliation{Landau Institute for Theoretical Physics,
Russian Academy of Sciences,
Chernogolovka, Moscow region, 142432 Russia \linebreak
and   \linebreak
Moscow Institute of Physics and Technology, Department
of Theoretical Physics, Dolgoprudnyj, Moskow region,
141707 Russia}

\begin{abstract}
The unitarity of the 4D lattice theory of gravity in the case of the Minkowski signature is proved.
The proof is valid only for lattices that conserve the number of degrees of freedom during time evolution.
The Euclidean signature and the Minkowski signature are related by the deformation of the integration contours of dynamic variables in a discrete lattice functional integral. It is important that the result is obtained directly on the lattice. Since the studied lattice theory of gravity in the long-wave limit transforms into the well-known Einstein-Cartan-Palatini theory, the obtained result means that this lattice theory of gravity has the right to be considered as a discrete regularization of the generally accepted continuous physical theory of gravity.
\end{abstract}


\maketitle

\section{Introduction}

The idea of discretizing space-time to describe the theory of gravity at a fundamental level
was first formulated by T. Regge in \cite{regge1961general}. In this work, the discretization of space-time was carried out using a simplicial complex. Each 1-simplex was assigned its length, so that all the sizes of each 2-simplex (triangle) were fixed. The lengths of the three sides of each triangle satisfied the triangle inequality. In this way, the geometry of the entire complex was fixed. A detailed description of the Regge calculus is given in \cite{friedberg1986derivation}. An approach to discrete geometry similar to Regge calculus can also be found in \cite{cheeger1984w,pinkall1996discrete}.

Despite the obvious elegance of the Regge calculus, this theory turns out to be very inconvenient when moving to quantum theory. Indeed, the independent variables determining the Regge action are the lengths of one-dimensional simplices, subject to a large number of constraints, namely, triangle inequalities. Moreover, the introduction of Dirac fields into the theory creates a new difficulty, consisting in the absence of explicit orthonormal bases. Perhaps this is why the version of discrete gravity based on the so-called B-F theory soon gained more intensive development. The B-F theory is developed on the basis of the action of the theory of gravity in the form of Palatini (see below (\ref{Long_Wav_Grav_Act})). The most characteristic property of B-F theory is that the curvature tensor of the 2-form is zero. However, while B-F theory does describe gravity in three-dimensional space, it does not do so in higher-dimensional spaces.
For example, in a four-dimensional space with action (\ref{Long_Wav_Grav_Act})) the curvature tensor
is not equal to zero. But this is not the only difficulty of the theory: the introduction of matter is also inconvenient. A detailed description of the discrete quantum theory of gravity based on the B-F formalism can be found in \cite{regge2000discrete,baez1999introduction,reisenberger1997lattice,iwasaki1999surface}.

In contrast to the multidimensional case, significant computational progress has been made in two-dimensional discrete quantum gravity \cite{david1985model,boulatov1986analytical}.

Later, more quantizable models of discrete gravity emerged \cite{vergeles2024alternative,vergeles2006one,vergeles2015wilson,vergeles2017note,vergeles2017fermion,
vergeles2021note,vergeles2021domain,vergeles2024phase,diakonov2011towards,vladimirov2012phase,
vladimirov2014diffeomorphism}. These models are characterized by the fact that in the naive long-wave limit they are transformed into the well-known Einstein-Cartan-Palatini continuum theory. By Einstein-Cartan-Palatini theory we mean a theory with the Hilbert-Einstein action, plus the action of the Dirac field minimally coupled to gravity, plus a contribution to the action from the cosmological constant, all of which are considered in Cartan-Palatini form.

The Einstein-Cartan-Palatini action is invariant under diffeomorphisms. The following fact is interesting here. In the lattice theory of gravity \cite{vergeles2024alternative,vergeles2006one,vergeles2015wilson,vergeles2017note,vergeles2017fermion,
vergeles2021note,vergeles2021domain,vergeles2024phase,diakonov2011towards,vladimirov2012phase,
vladimirov2014diffeomorphism} local coordinates are completely absent; they appear only when passing to the long-wave limit, and they can be introduced arbitrarily. The long-wave limit is nothing more than the expansion of the original action in terms of a small parameter $\xi\sim l_P/\lambda\ll1$ that is independent of local coordinates. Here $l_P$ is the Planck scale, and $\lambda$ is the physical scale. It is natural to assume that $l_P$ is the characteristic scale of one 1-simplex of the lattice, and $\lambda\sim n l_P$, where $n\gg1$ is the number of 1-simplices of the graph that fits into the wavelength. From the above it is clear that the parameter $\xi$ does not depend on the local coordinates, which are arbitrary. Therefore, when expanding the lattice action in the parameter $\xi$, each term of a given degree relative to $\xi$ is invariant under diffeomorphisms. In this expansion, the main term is the Einstein-Cartan-Palatini action whose invariance with respect to diffeomorphisms is established independently. The remaining terms, which quickly die out according to a power law in the long-wave limit, are also invariants with respect to diffeomorphisms (see Section II).

The following problem of quantization of generally covariant theories should be pointed out. Such theories represent degenerate dynamical systems with constraints. To construct a unitary and covariant S-matrix, it is necessary to prove: 1) the gauge-independence of the covariant S-matrix and 2) its unitarity at least in one (non relativistic) gauge. If both requirements can be fulfilled, then the unitarity is guaranteed also in a relativistic gauge, and the suggested Feynman rules are correct.
To construct a unitary S-matrix, the Hamiltonian and all constraints are expressed in terms of the chosen independent physical canonical variables, the gauge is fixed, the functional measure is determined, and the transition amplitude is written out in the form of a functional integral \cite{dirac1950generalized}. Thus the second requirement is fulfilled. To satisfy the first requirement, many theories (for example Yang-Mills theory) use the Faddeev-Popov trick \cite{faddeev1974covariant}.
However, in the case of theories that are invariant under diffeomorphisms (for example, the theory of gravity), the Faddeev-Popov technology is fundamentally unsuitable.  The reason for this is that, under general coordinate transformations in chronologically ordered averages, the order of the operators, generally speaking, changes. These difficulties were overcome in the works of Fradkin and Fradkin-Vilkovisky
\cite{fradkin1975quantization}. Although Fradkin-Vilkovisky technology is extremely complex, it fundamentally solves the problem.

From the above it follows that the Einstein-Cartan-Palatini quantum theory allows the construction of a unitary and gauge-invariant S-matrix. However, this model is only the main term in the expansion of the lattice theory under study in the parameter $\xi$. An important question arises: do the following terms of this expansion violate the unitarity of the theory? If this were so, then the lattice model of gravity that we are studying could not be considered as a model that even slightly describes the real world.

We now formulate the result of the present work: the evolution of the quantum model of discrete gravity formulated in Section II is a unitary transformation. Let us emphasize that this result was obtained directly in lattice theory and, of course, for the Minkovsky signature. Note also that, unlike the continuum theory of gravity, the lattice theory of gravity has no local coordinates at all, since neither Cartesian spaces nor manifolds are used in defining abstract simplicial complexes \cite{pontryagin1976basics}. We will see that this circumstance radically simplifies the proof of unitarity of lattice theory.

The work is organized as follows.

In Section II we define the lattice gravity model we study in two related versions: the Euclidean signature and the Minkowski signature. These two versions of the theory transform into each other by deformation in the complex plane of the integration contours over the dynamic variables in the functional integral, representing either the partition function or the transition amplitude.

In Section III, a proof of the unitarity of the transition amplitude in lattice theory is given.

The Appendix provides some information from the field of combinatorial topology necessary for proving unitarity.

\section{Definition of Lattice Theory of Gravity}

In the works of the author \cite{vergeles2024alternative,vergeles2006one,vergeles2015wilson,vergeles2017note,vergeles2017fermion,
vergeles2021note,vergeles2021domain,vergeles2024phase} the lattice theory of gravity coupled to Dirac fields was studied. Let us give a brief definition of it.

Consider the orientable 4-dimensional simplicial complex $\mK$. We recommend the book
\cite{pontryagin1976basics}, \S\S 2.4 for an introduction to the definition of abstract simplicial complexes. Suppose that each of its 4-simplexes belongs to a subcomplex of ${\mK}'\in\mK$ that has a geometric realization in $\R^4$ without cavities. The vertices are denoted by $a_{\cV}$, the indices ${\cV}$ and ${\cW}$ numerate the vertices and 4-simplices $s^4_{\cW}$, respectively. It is necessary to use local enumeration of vertices belonging to a given 4-simplex: all 5 vertices of the 4-simplex $s^4_{\cW}$ are numbered as $a_{{\cV}_{({\cW})i}}$, $i=1,2,3,4,5$. In what follows, notations with an additional subscript $({\cW})$ indicate that the corresponding object belongs to the 4-simplex $s^4_{\cW}$.
Let us denote
$\varepsilon_{{\cV}_{({\cW})1}{\cV}_{({\cW})2}{\cV}_{({\cW})3}{\cV}_{({\cW})4}{\cV}_{({\cW})5}}=\pm 1$
the Levi-Civita symbol. The upper (lower) sign depends on the orientation of the 4-simplex
$s^4_{\cW}=a_{{\cV}_{({\cW})1}}a_{{\cV}_{({\cW})2}}
a_{{\cV}_{({\cW})3}}a_{{\cV}_{({\cW})4}}a_{{\cV}_{({\cW})5}}$.

\subsection{Euclidean signature}

Element of the compact group $\Spin(4)$ and element of the Clifford algebra
\begin{gather}
\Omega_{{\cV}_1{\cV}_2}=\Omega^{-1}_{{\cV}_2{\cV}_1}=\exp\left(
\omega_{{\cV}_1{\cV}_2}\right)
\nonumber \\
=\exp\left(\frac{1}{2}\sigma^{ab}
\omega^{ab}_{{\cV}_1{\cV}_2}\right)\in\Spin(4),  \quad
\sigma^{ab}\equiv\frac14[\gamma^a\gamma^b],
\nonumber \\
\gamma^a\gamma^b+\gamma^b\gamma^a=2\delta^{ab},  \quad a=1,2,3,4, \quad
\gamma^5\equiv\gamma^1\gamma^2\gamma^3\gamma^4,
\nonumber \\
\hat{e}_{{\cV}_1{\cV}_2}\equiv e^a_{{\cV}_1{\cV}_2}\gamma^a\equiv
-\Omega_{{\cV}_1{\cV}_2}\hat{e}_{{\cV}_2{\cV}_1}\Omega_{{\cV}_1{\cV}_2}^{-1},
\nonumber \\
|e_{{\cV}_1{\cV}_2}|<1, \quad |e_{{\cV}_1{\cV}_2}|\equiv\sqrt{\sum_a(e^a_{{\cV}_1{\cV}_2})^2}
\label{Variables_Grav}
\end{gather}
are defined on each oriented 1-simplex $a_{{\cV}_1}a_{{\cV}_2}$. By assumption, the set of variables $\{\Omega,\,\hat{e}\}$ is a set of independent bosonic dynamical variables. Fermionic degrees of freedom (Dirac spinors) are defined at the vertices of the complex:
\begin{gather}
\Psi^{\dag}_{\cV}, \quad \Psi_{\cV}.
\label{Variables_Ferm}
\end{gather}
The set of variables $\{\Psi^{\dag},\,\Psi\}$ are mutually independent, and the spinors $\Psi^{\dag}_{\cV}$ and $\Psi_{\cV}$ are in mutual involution (or anti-involution) with respect to the operation of Hermitian conjugation.

Let's consider a model with action
\begin{gather}
\mA=\mA_g+\mA_{\Psi}+\mA_{\Lambda_0}.
\label{Action_4D}
\end{gather}
Here $\mA_g$ and $\mA_{\Psi}$ are the actions of pure gravity and the Dirac field, respectively:
\begin{widetext}
\begin{gather}
\mA_g=-\frac{1}{5!\cdot2\cdot l_P^2}\sum_{\cW}\sum_{\sigma}
\varepsilon_{\sigma({\cV}_{({\cW})1})\sigma({\cV}_{({\cW})2})
\sigma({\cV}_{({\cW})3})\sigma({\cV}_{({\cW})4})\sigma({\cV}_{({\cW})5})}
\nonumber \\
\times\tr\gamma^5\bigg\{
\Omega_{\sigma({\cV}_{({\cW})5})\sigma({\cV}_{({\cW})1})}
\Omega_{\sigma({\cV}_{({\cW})1})\sigma({\cV}_{({\cW})2})}\Omega_{\sigma({\cV}_{({\cW})2})\sigma({\cV}_{({\cW})5})}
\hat{e}_{\sigma({\cV}_{({\cW})5})\sigma({\cV}_{({\cW})3})}
\hat{e}_{\sigma({\cV}_{({\cW})5})\sigma({\cV}_{({\cW})4})}\bigg\}.
\label{Latt_Action_Grav}
\end{gather}
\begin{gather}
\mA_{\Psi}=\frac{1}{5\cdot24^2}\sum_{\cW}\sum_{\sigma}
\varepsilon_{\sigma({\cV}_{({\cW})1})\sigma({\cV}_{({\cW})2})
\sigma({\cV}_{({\cW})3})\sigma({\cV}_{({\cW})4})\sigma({\cV}_{({\cW})5})}
\nonumber \\
\times\tr\gamma^5\bigg\{ \hat{\Theta}_{\sigma({\cV}_{({\cW})5})\sigma({\cV}_{({\cW})1})}
\hat{e}_{\sigma({\cV}_{({\cW})5})\sigma({\cV}_{({\cW})2})}
\hat{e}_{\sigma({\cV}_{({\cW})5})\sigma({\cV}_{({\cW})3})}
\hat{e}_{\sigma({\cV}_{({\cW})5})\sigma({\cV}_{({\cW})4})}\bigg\},
\label{Latt_Action_Ferm}
\end{gather}
Each $\sigma$ is one of 5! permutations of vertices ${\cV}_{({\cW})i}\longrightarrow\sigma(
{\cV}_{({\cW})i})$.
\begin{gather}
\hat{\Theta}_{{\cV}_1{\cV}_2}
\equiv\Theta^a_{{\cV}_1{\cV}_2}\gamma^a=\hat{\Theta}_{{\cV}_1{\cV}_2}^{\dag},  \quad
\Theta^a_{{\cV}_1{\cV}_2}=\frac{i}{2}\left(\Psi^{\dag}_{{\cV}_1}\gamma^a
\Omega_{{\cV}_1{\cV}_2}\Psi_{{\cV}_2}-
\Psi^{\dag}_{{\cV}_2}\Omega_{{\cV}_2{\cV}_1}\gamma^a\Psi_{{\cV}_1}\right).
\label{Dirac_Form}
\end{gather}
It can be verified that (compare with (\ref{Variables_Grav}))
\begin{gather}
\hat{\Theta}_{{\cV}_1{\cV}_2}
\equiv-\Omega_{{\cV}_1{\cV}_2}\hat{\Theta}_{{\cV}_2{\cV}_1}
\Omega_{{\cV}_1{\cV}_2}^{-1}.
\label{Dir_Bil_Form_Trans}
\end{gather}
The contribution to the lattice action from the cosmological constant is
\begin{gather}
\mA_{\Lambda_0}=-\frac{1}{5!\cdot12}\cdot\Lambda_0\cdot\varepsilon_{abcd}\sum_{\cW}\sum_{\sigma}
\varepsilon_{\sigma({\cV}_{({\cW})1})\sigma({\cV}_{({\cW})2})
\sigma({\cV}_{({\cW})3})\sigma({\cV}_{({\cW})4})\sigma({\cV}_{({\cW})5})}
\nonumber \\
\times e^a_{\sigma({\cV}_{({\cW})5})\sigma({\cV}_{({\cW})1})}
e^b_{\sigma({\cV}_{({\cW})5})\sigma({\cV}_{({\cW})2})}
e^c_{\sigma({\cV}_{({\cW})5})\sigma({\cV}_{({\cW})3})}
e^d_{\sigma({\cV}_{({\cW})5})\sigma({\cV}_{({\cW})4})}.
\label{Latt_Action_Lambda}
\end{gather}
\end{widetext}
The partition function is represented by an integral
\begin{gather}
Z=\prod_{1-\mbox{simplices}}\int_{|e_{{\cV}_1{\cV}_2}|<1}\prod_a\d e^a_{{\cV}_1{\cV}_2}
\int\d\mu\{\Omega_{{\cV}_1{\cV}_2}\}
\nonumber \\
\times\prod_{\cV}\int\d\Psi^{\dag}_{\cV}\d\Psi_{\cV}\exp(\mA).
\label{Partition_function}
\end{gather}

The action (\ref{Action_4D}), as well as the integral (\ref{Partition_function}), are invariant under gauge transformations
\begin{gather}
\tilde{\Omega}_{{\cV}_1{\cV}_2}
=S_{{\cV}_1}\Omega_{{\cV}_1{\cV}_2}S^{-1}_{{\cV}_2}, \quad
\tilde{\hat{e}}_{{\cV}_1{\cV}_2}=S_{{\cV}_1}\,\hat{e}_{{\cV}_1{\cV}_2}\,S^{-1}_{{\cV}_1},
\nonumber \\
\tilde{\Psi}_{\cV}=S_{\cV}\Psi_{\cV}, \quad \tilde{\Psi^{\dag}}_{\cV}=\Psi_{\cV}^{\dag}S_{\cV}^{-1}, \quad  S_{\cV}\in\Spin(4).
\label{Gauge_Trans}
\end{gather}
Verification of this fact is made easier by using the relation (compare with the relation
for $\hat{e}_{{\cV}_1{\cV}_2}$ in (\ref{Gauge_Trans}))
\begin{gather}
\tilde{\hat{\Theta}}_{{\cV}_1{\cV}_2}=S_{{\cV}_1}\hat{\Theta}_{{\cV}_1{\cV}_2}S^{-1}_{{\cV}_1},
\label{Teta_Gauge_Trans}
\end{gather}
which follows directly from (\ref{Gauge_Trans}).

The lattice model under consideration is invariant with respect to the global discrete $\Z_2$-symmetry, which is an analogue of the combined $PT$-symmetry. Let $\hat{\cal U}_{PT}$ denote the operator of this transformation. Then the transformed dynamic variables are expressed through the original variables as follows:
\begin{gather}
\hat{\cal U}_{PT}^{-1}\Psi_{\cV}\hat{\cal U}_{PT}=U_{PT}\left(\Psi^{\dag}_{\cV}\right)^t,
\nonumber \\
\hat{\cal U}_{PT}^{-1}\Psi^{{\dag}}_{\cV}\hat{\cal U}_{PT}=-\left(\Psi_{\cV}\right)^tU^{-1}_{PT}, \quad U_{PT}=i\gamma^1\gamma^3
\nonumber \\
\hat{\cal U}_{PT}^{-1}e^a_{{\cV}_1{\cV}_2}\hat{\cal U}_{PT}=-e^{a}_{{\cV}_1{\cV}_2}, \quad
\hat{\cal U}_{PT}^{-1}\omega^{ab}_{{\cV}_1{\cV}_2}\hat{\cal U}_{PT}=\omega^{ab}_{{\cV}_1{\cV}_2}.
\label{PT_transform}
\end{gather}
Here the superscript $"t"$ denotes the transposition of the Dirac matrices and spinors. We have:
\begin{gather}
U^{-1}_{PT}\gamma^aU_{PT}=(\gamma^a)^t, \quad
U^{-1}_{PT}\sigma^{ab}U_{PT}=-(\sigma^{ab})^t.
\label{PT_trans_Dir_Alg}
\end{gather}
From (\ref{PT_transform}) and (\ref{PT_trans_Dir_Alg}) it follows that
\begin{gather}
 U^{-1}_{PT}\Omega_{{\cV}_1{\cV}_2}U_{PT}=\left(\Omega_{{\cV}_2{\cV}_1}\right)^t,
\label{PT_trans_Conn}
\end{gather}
\begin{gather}
\hat{\cal U}_{PT}^{-1}\Theta^a_{{\cV}_1{\cV}_2}\hat{\cal U}_{PT}=-\Theta^a_{{\cV}_1{\cV}_2}.
\label{PT_trans_Ferm_Bil_Fjrm}
\end{gather}
We see that the action (\ref{Action_4D}) is invariant under the transformations (\ref{PT_transform})-(\ref{PT_trans_Ferm_Bil_Fjrm}).

\subsection{Minkowski signature}

Let's pass on to the Minkowski signature.
Further in this section, all lattice variables
in the case of the Euclidean signature are provided with a prime. For field variables in the case of the Minkowski signature, the old notations are used.

For the specified transformation of the action, the following deformations of the integration contours in the integral (\ref{Partition_function}) are necessary:
\begin{gather}
{\omega'}_{{\cV}_1{\cV}_2}^{4\alpha}= i\omega^{0\alpha}_{{\cV}_1{\cV}_2}, \quad
{\omega'}_{{\cV}_1{\cV}_2}^{\alpha\beta}=-\omega_{{\cV}_1{\cV}_2}^{\alpha\beta},
\nonumber \\
{e'}^4_{{\cV}_1{\cV}_2}= e^0_{{\cV}_1{\cV}_2}, \quad {e'}^{\alpha}_{{\cV}_1{\cV}_2}= ie^{\alpha}_{{\cV}_1{\cV}_2}.
\label{Variables_Trans_Mink}
\end{gather}
The variables $\omega^{ab}_{{\cW}ij}$, $e^a_{{\cW}ij}$ in the Minkowski signature are real, and their indices take the values
\begin{gather}
a,\,b\ldots=0,1,2,3, \quad \alpha,\,\beta,\ldots=1,2,3.
\label{Mink_Ind}
\end{gather}
In the orthonormal basis (ONB) the metric tensor $\eta^{ab}=\diag(1,\,-1,\,-1,\,-1)$.
Dirac matrices are transformed as
\begin{gather}
{\gamma'}^4=\gamma^0, \quad {\gamma'}^{\alpha}= i\gamma^{\alpha}, \quad
{\gamma'}^5=\gamma^5=i\gamma^0\gamma^1\gamma^2\gamma^3.
\label{Mink_Dirac_matr}
\end{gather}
Thus, for spin matrices $\sigma^{ab}=(1/4)[\gamma^a,\,\gamma^b]$ we have:
\begin{gather}
{\sigma'}^{4\alpha}= i\sigma^{0\alpha}, \quad {\sigma'}^{\alpha\beta}=-\sigma^{\alpha\beta}.
\label{Mink_Spin_Matr}
\end{gather}
When passing to the Minkowski signature, the Dirac spinors are transformed as follows:
\begin{gather}
\Psi'_{\cV}=\Psi_{\cV}, \quad  {\Psi'}^{\dag}_{\cV}=\Psi_{\cV}^{\dag}\gamma^0=\overline{\Psi}_{\cV}.
\label{Mink_Dir}
\end{gather}
Using (\ref{Variables_Trans_Mink})-(\ref{Mink_Dir}) we find:
\begin{gather}
{\omega'}_{{\cV}_1{\cV}_2}=\frac12\omega^{ab}_{{\cV}_1{\cV}_2}\,\sigma_{ab}
\equiv\omega_{{\cV}_1{\cV}_2},
\nonumber \\
{\hat{e}'}_{{\cV}_1{\cV}_2}=\gamma_ae^a_{{\cV}_1{\cV}_2}\equiv\hat{e}_{{\cV}_1{\cV}_2},
\nonumber \\
{\Omega'}_{{\cV}_1{\cV}_2}=\exp\left(\frac12{\omega'}_{{\cV}_1{\cV}_2}^{ab}{\sigma'}^{ab}\right)
\nonumber \\
=\exp\left(\frac12\omega_{{\cV}_1{\cV}_2}^{ab}
\sigma_{ab}\right)\equiv\Omega_{{\cV}_1{\cV}_2}\in\Spin(3,1),
\nonumber \\
\hat{\Theta}'_{{\cV}_1{\cV}_2}=\gamma_a\Theta^a_{{\cV}_1{\cV}_2}\equiv\hat{\Theta}_{{\cV}_1{\cV}_2},
\nonumber \\
\Theta^a_{{\cV}_1{\cV}_2}=\frac{i}{2}\left(\overline{\Psi}_{{\cV}_1}\gamma^a
\Omega_{{\cV}_1{\cV}_2}\Psi_{{\cV}_2}-
\overline{\Psi}_{{\cV}_2}\Omega_{{\cV}_2{\cV}_1}\gamma^a\Psi_{{\cV}_1}\right)
\nonumber \\
=(\Theta^a_{{\cV}_1{\cV}_2})^{\dag}.
\label{Mink_5}
\end{gather}
In the Minkowski signature the action still has the form (\ref{Action_4D}), but it is a functional of the unprimed variables from the right-hand sides of the equations (\ref{Mink_5}) and is purely imaginary. This is a result of the deformation of the integration contours according to (\ref{Variables_Trans_Mink}) and the substitutions (\ref{Mink_Dir}). Let us prove this fact.

Let us write out the elementary term from the action (\ref{Latt_Action_Grav}) with an accuracy of up to a real factor, and let us represent the Euclidean primed variables in the form (\ref{Variables_Trans_Mink}).
Thus we go to the Minkowski signature:
\begin{gather}
\varepsilon_{\cV_1\cV_2\cV_3\cV_4\cV_5}\tr\gamma^5\Omega'_{\cV_5\cV_1}\Omega'_{\cV_1\cV_2}\Omega'_{\cV_2\cV_5}
\hat{e}'_{\cV_5\cV_3}\hat{e}'_{\cV_5\cV_4}
\nonumber \\
=\varepsilon_{\cV_1\cV_2\cV_3\cV_4\cV_5}\tr\gamma^5\Omega_{\cV_5\cV_1}\Omega_{\cV_1\cV_2}\Omega_{\cV_2\cV_5}
\hat{e}_{\cV_5\cV_3}\hat{e}_{\cV_5\cV_4}\equiv i{\cal A}.
\label{Term_Grav}
\end{gather}
The first equality in (\ref{Term_Grav}) holds due to the equalities (\ref{Mink_5}). Let us show that ${\cal A}$ is a real quantity. To do this, we perform the Hermitian conjugation operation of the quantity (\ref{Term_Grav}):
\begin{gather}
-i{\cal A}^{\dag}=\varepsilon_{\cV_1\cV_2\cV_3\cV_4\cV_5}\tr\gamma^5\hat{e}_{\cV_5\cV_4}^{\dag}
\hat{e}_{\cV_5\cV_3}^{\dag}\Omega_{\cV_2\cV_5}^{\dag}\Omega_{\cV_1\cV_2}^{\dag}\Omega_{\cV_5\cV_1}^{\dag}.
\label{Herm_Conj_Grav}
\end{gather}
Now let's take into account the equalities
\begin{gather}
\hat{e}_{\cV_i\cV_j}^{\dag}=\gamma^0\hat{e}_{\cV_i\cV_j}\gamma^0, \quad
\Omega_{\cV_i\cV_j}^{\dag}=\gamma^0\Omega_{\cV_j\cV_i}\gamma^0,
\nonumber \\
\gamma^0\gamma^5\gamma^0=-\gamma^5
\label{Equalities}
\end{gather}
and use them to transform the value (\ref{Herm_Conj_Grav}):
\begin{gather}
-i{\cal A}^{\dag}=-\varepsilon_{\cV_1\cV_2\cV_3\cV_4\cV_5}\tr\gamma^5\hat{e}_{\cV_5\cV_4}
\hat{e}_{\cV_5\cV_3}\Omega_{\cV_5\cV_2}\Omega_{\cV_2\cV_1}\Omega_{\cV_1\cV_5}
\nonumber \\
=-\varepsilon_{\cV_1\cV_2\cV_3\cV_4\cV_5}\tr\gamma^5\Omega_{\cV_5\cV_1}\Omega_{\cV_1\cV_2}\Omega_{\cV_2\cV_5}
\hat{e}_{\cV_5\cV_3}\hat{e}_{\cV_5\cV_4}
\nonumber \\
\equiv -i{\cal A}.
\label{Herm_Conj_Grav_II}
\end{gather}
Here we have taken advantage of the fact that simultaneous permutations $\cV_1\leftrightarrow \cV_2$ and
$\cV_3\leftrightarrow \cV_4$ leave the symbol $\varepsilon_{\cV_1\cV_2\cV_3\cV_4\cV_5}$ invariant.
From (\ref{Herm_Conj_Grav_II}) it is clear that the quantity ${\cal A}$ is Hermitian, and therefore the contribution to the action (\ref{Latt_Action_Grav}) is an anti-Hermitian quantity in the Minkowski signature.

The fact that the contributions to the action (\ref{Latt_Action_Ferm}) and (\ref{Latt_Action_Lambda}) become anti-Hermitian in the Minkowski signature is already evident from the equality
$\tr\gamma^5\gamma_a\gamma_b\gamma_c\gamma_d=4i\varepsilon_{abcd}$.

We see that the full action (\ref{Action_4D}) becomes anti-Hermitian in the Minkowski signature:
\begin{gather}
e^{\mA'}=e^{i\mA}, \quad \mA^{\dag}=\mA.
\label{Minkowski signature}
\end{gather}

It should be noted that the Hermitian property of the expression $\mA$ in (\ref{Minkowski signature}) is formal, reflecting only the formal application of the operator ${}^{\dag}$ to all terms in the action $\mA$.
To actually determine whether an operator is Hermitian, it is necessary to have a definition of the scalar product in the vector space where the operator acts. This problem is solved further.

The transition to the long-wave limit is possible for such field configurations that change slowly enough during transitions from simplex to simplex, i.e. during small or significant displacements along the lattice. In our theory, it is at the stage of transition to the long-wave limit that the need arises to introduce local coordinates. Local coordinates $x^{\mu}$ ($\mu=0,1,2,3$) are markers of the lattice vertices. Without going into details, we write out the long-wave limit of action (\ref{Action_4D}) in the Minkowski signature
\cite{vergeles2006one,vergeles2015wilson,vergeles2017note,vergeles2017fermion,
vergeles2021note,vergeles2021domain,vergeles2024phase,vergeles2024alternative}:
\begin{gather}
{\mA'}_g\longrightarrow i\mA_g,  \quad \mA_g=-\frac{1}{4\,l^2_P}\varepsilon_{abcd}\int\mR^{ab}\wedge e^c\wedge e^d,
\nonumber \\
\frac12\sigma_{ab}\mR^{ab}=\frac12\sigma_{ab}\mR^{ab}_{\mu\nu}\d x^{\mu}\wedge\d x^{\nu}
\nonumber \\
=\big(\partial_{\mu}\omega_{\nu}-\partial_{\nu}\omega_{\mu}+
[\omega_{\mu},\,\omega_{\nu}\,]\big)\d x^{\mu}\wedge\d x^{\nu},
\label{Long_Wav_Grav_Act}
\end{gather}
\begin{gather}
{\mA'}_{\Psi}\longrightarrow i\mA_{\Psi}, \quad \mA_{\Psi}=\frac16\varepsilon_{abcd}\int\Theta^a\wedge e^b\wedge e^c\wedge e^d,
\nonumber \\
\Theta^a=\frac{i}{2}\left[\overline{\Psi}\gamma^a{\cal D}_{\mu}\,\Psi-
\left(\overline{{\cal D}_{\mu}\,\Psi}\right)\gamma^a\Psi\right]\d x^{\mu},
\nonumber \\
{\cal D}_{\mu}=\left(\partial_{\mu}+\omega_{\mu}\right),
\label{Long_Wav_Dir_Act}
\end{gather}
\begin{gather}
\mA'_{\Lambda_0}\longrightarrow i\mA_{\Lambda_0}, \quad
\mA_{\Lambda_0}=-2\Lambda_0\int e^0\wedge e^1\wedge e^2\wedge e^3.
\label{Long_Wav_Lambda_Act}
\end{gather}
Here 1-forms
\begin{gather}
\omega_{\mu}^{ab}\d x^{\mu},  \quad  e^a_{\mu}\d x^{\mu}
\label{1-forms}
\end{gather}
are the long-wave limits of the lattice variables (\ref{Variables_Trans_Mink}).

All other terms in such a transition will contain additional factor $\xi\sim(l_P/\lambda)\longrightarrow0$ to the positive power, and therefore they are omitted ($\lambda$ is the characteristic wavelength of the physical subsystem).

The action (\ref{Long_Wav_Grav_Act})-(\ref{Long_Wav_Lambda_Act}) is the Einstein-Cartan-Palatini action minimally coupled to the Dirac field.

\section{Time evolution of states in the lattice theory of gravity.}

Next we study lattice theory in Minkowski signature.

Let $\Sigma$ denote the 3D simplicial complex on which the wave function (WF) of lattice gravity is defined.
The WF is a functional of bosonic variables (\ref{Variables_Grav}) defined on 1-simplices and fermionic variables (\ref{Variables_Ferm}) defined at the vertices, and it is invariant under gauge transformations (\ref{Gauge_Trans}), but with $S_{\cV}\in\Spin(3,1)$. It is assumed that the 3D complex $\Sigma$ models a spatial lattice. Therefore, bosonic variables (\ref{Variables_Grav}) defined on the complex $\Sigma$ are dynamic variables.

Note that in the long-wave limit, the 3D  complex $\Sigma$ transforms into a spatially similar curved hypersurface $\Sigma$ in a 4D manifold $\mK$. Let $x^{\mu}$ be local coordinates in $\mK$. We can assume that the hypersurface $\Sigma$ is given by the equation $x^0=0$. Then the  coordinates $x^i$, $i=1,2,3$, are local coordinates on the hypersurface $\Sigma$. It is known that the set of bosonic dynamic variables of the system (\ref{Long_Wav_Grav_Act})-(\ref{Long_Wav_Lambda_Act}) is determined by the restriction of 1-forms (\ref{1-forms}) on $\Sigma$. Therefore, the bosonic dynamical variables are the fields $\omega_i^{ab}$ and  $e^a_i$, and the fields $\omega_0^{ab}$ and  $e^a_0$ play the role of Lagrange multipliers for the constraints imposed on the fields $\omega_i^{ab}$ and  $e^a_i$. In the functional integral for the transition amplitude, the effective action and the functional measure are very complicated \cite{fradkin1975quantization}: The effective action contains, in addition to the original contribution (\ref{Long_Wav_Grav_Act})-(\ref{Long_Wav_Lambda_Act}), also contributions fixing all gauges and the contribution of the additional fermion fields (ghosts). The latter contribution is nonlinear relative to the ghost fields.

From the above it follows that in lattice theory the WF is defined on physical boson degrees of freedom (\ref{Variables_Grav}), which are analogous to fields $\omega_i^{ab}$ and  $e^a_i$ in continuum theory.

\subsection{Local time evolution operator. First example}

The mechanism of the evolution of the WF in the lattice theory, described below, is radically different from the corresponding mechanism in the continuum theory. These differences boil down to two points:

1) In the continuum theory, the functional integral determines the transition amplitude immediately for the entire WF of the system defined on the hypersurface $\Sigma$. On the contrary, in lattice theory there is the possibility of defining a local transition amplitude or transfer matrix that affects the evolution of degrees of freedom only on those lattice simplices that contain one (arbitrary) vertex. This method is implemented here.

2) The second difference is that in the lattice theory there are no local coordinates. This fact radically simplifies the construction of the integration measure and secondary constraints. The secondary constraints are reduced to fixing the gauge, which is done without changing the original gauge-invariant measure. The effective action remains unchanged, it coincides with the original action in the Minkowski signature.

Let us proceed to constructing the local transfer matrix and proving its unitarity.

Let us begin by describing the 4-complex that defines the local transfer matrix that translates the WF in time locally near the vertex $a_p\subset\Sigma$. As a result, the vertex $a_p$ and its nearest simplices advance in time to the point $a_{p'}$, which is outside the complex $\Sigma$ ahead in time. The 3D complex $\Sigma$ is transformed into a 3D complex $\Sigma'\supset a_{p'}$, on which a new WF is defined. Let us denote the bounded 4D-complex that transforms $\Sigma$ into $\Sigma'$ as $K^4_{pp'}$. The described action is schematically depicted in Fig. 1 in the case of 1D complexes $\Sigma$ and $\Sigma'$.

\begin{figure}[t]
\includegraphics[width=\linewidth]{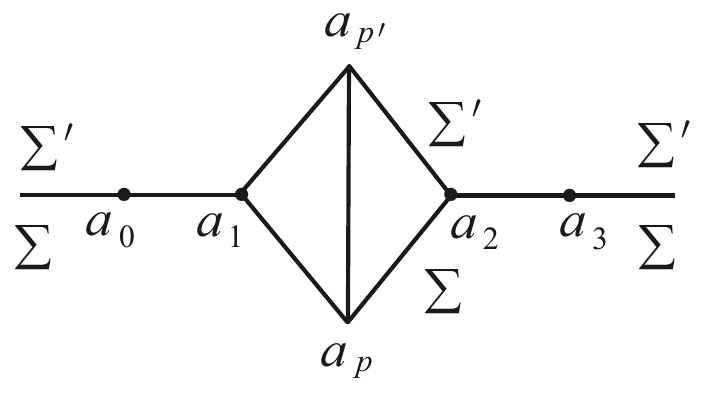}
\caption{The 1-complex $\Sigma$ is the lower broken line, and near the vertex $a_p$ it contains 1-simplices $(a_1a_p)$, $(a_pa_2)$. The 1-complex $\Sigma'$ differs from $\Sigma$ only in that it contains  1-simplices $(a_1a_{p'})$, $(a_{p'}a_2)$  instead of 1-simplices $(a_1a_p)$, $(a_pa_2)$. Here the 2-complex $K^2_{pp'}$ is the union of two 2-simplices: $K^2_{pp'}=(a_1a_pa_{p'})\cup(a_2a_p'a_{p})$.}
\label{Fig_1}
\end{figure}

Let an arbitrary vertex $a_p\in\Sigma$, and denote by $K_p^3\subset\Sigma$ the subcomplex consisting of all 3-simplices $s^3_{1p},\ldots,s^3_{np}$ containing $a_p$: $a_p\subset s^3_{ip}$, $i=1,\ldots,n$. Thus
$s^3_{ip}=(a_{i1}a_{i2}a_{i3}a_p)$, where $a_{i1},a_{i2},a_{i3},a_p$ are the four vertices of the simplex $s^3_{ip}$, $K^3_p=\cup_i^ns^3_{ip}$. Of course, each of the vertices $a_{i\rho}\in s^3_{ip}$, $\rho=1,2,3$,  also belongs to the adjacent simplices $s^3_{i'p}$. The boundary of a subcomplex $K^3_p$ is as follows: $\partial K_p^3=\cup_i^n
(a_{i1}a_{i2}a_{i3})$.

Let the 0-simplex $a_{p'}$ do not belong to the complex $\Sigma$: $a_{p'}\notin\Sigma$. Let us consider the set of 3-simplices $s^3_{ip'}=(a_{i1}a_{i2}a_{i3}a_{p'})$, as well as the 3-complex $K^3_{p'}=\cup_i^ns^3_{ip'}$. We have:
\begin{gather}
\partial K^3_p=\partial K^3_{p'}=S^2.
\label{Boundary}
\end{gather}
Here $S^2$ is a two-dimensional sphere.

Let us introduce into consideration 4-simplices
\begin{gather}
s^4_{ipp'}=(a_{i1}a_{i2}a_{i3}a_pa_{p'}), \quad i=1,\ldots,n,
\label{s^4_ipp'}
\end{gather}
and the union of these 4-simplices:
\begin{gather}
K^4_{pp'}=\cup_i^ns^4_{ipp'}, \quad \partial K^4_{pp'}=K^3_p\cup K^3_{p'}.
\label{K^4pp'}
\end{gather}

It was indicated above that equalities $a_{i_1{\rho}_1}=a_{i_2{\rho}_2}$ hold for some different values of indices
$(i_1{\rho}_1)$ and $(i_2{\rho}_2)$. Therefore, from the set of vertices $\{a_{i\rho}\}$ we select $r$ different vertices $a_{\alpha}\subset\partial K^3_p$, $\alpha=1,\ldots,r$, that exhaust this set.

Let us denote the part of the action defined on $K^4_{pp'}$ as $\mA_{pp'}$. To translate a WF defined on a 3D complex $\Sigma$ from a vertex $a_p$ to a vertex $a_{p'}$ in the immediate vicinity of these vertices, the following operations must be performed: 1) The WF must be multiplied by the factor $\exp(i\mA_{pp'})$;
2) it is necessary to integrate over all variables defined at the vertex $a_p$ and 1-simplices
$(a_pa_{p'})$ and $(a_{\alpha}a_p)$, $\alpha=1,\ldots,r$. As a result, a WF is generated, defined on the 3D complex $\Sigma'$. Formally, the complex $\Sigma'$ differs from the complex $\Sigma$ by replacing vertex $a_p$ with vertex $a_{p'}$ everywhere. This means that all simplices $(a_{i1}a_{i2}a_{i3}a_p)\subset\Sigma$ are replaced by simplices $(a_{i1}a_{i2}a_{i3}a_{p'})\subset\Sigma'$. Since there is a one-to-one correspondence between all simplices of the complexes $\Sigma$ and $\Sigma'$, as a result of the described quantum evolution the number of degrees of freedom of the system is conserved. It seems to us that this is a very important property of the theory, and only such a theory can be unitary. The last statement is a hypothesis.

\subsection{Gauge fixing}

The gauge group $\Spin(3,1)$ is non-compact, and the described integral for the time evolution of the WF admits a finite number of gauge transformations that preserve both the measure and the integrand. Therefore, it is necessary to fix the gauge in the Minkowski signature. To solve this problem, some mathematical results concerning the homology theory of 1D simplicial complexes are needed. These results and the necessary terminology are given in the Appendix.

Let us consider a subcomplex of some abstract complex consisting of all its 1-simplices and 0-simplices and call this 1D subcomplex the one-dimensional skeleton or graph of the complex.
Let us choose a basis in the one-dimensional homology group of this graph, or more precisely, one representative for each basis element. Thus, the cycles $\mc_1,\ldots,\mc_C$ can be considered as a basis of the one-dimensional homology group of the graph. Each of these cycles is a closed broken contour consisting of 1-simplices connected in series. Of course, the set of cycles $\mc_1,\ldots,\mc_C$ representing the basis of a one-dimensional homology group is not unique. \\

{\it Statement 1}. To fix the gauge, it is necessary and sufficient to have on each basic element $\mc$ of the graph at least one independent variable $\Omega_{\cV_1\cV_2}$ related to a certain 1-simplex $a_1a_2\subset\mc$. The system contains the same number of independent variables $\Omega$ as the dimension of the one-dimensional homology group of the graph. All other variables $\Omega$ should be fixed (for example, are set equal to one). \\

Note that the choice of 1-simplices $(a_1a_2)\subset\mc$ on which the independent variables $\Omega_{\cV_1\cV_2}$ are defined is also ambiguous.

Bringing the set of variables $\{\Omega\}$ to the form specified in the Statement 1 is achieved by means of gauge transformations. Suppose that the graph contains 1-simplices that are not contained in any cycle. We will call the set of such 1-simplices a tree and denote it by $\mt$. Obviously, moving sequentially along the tree (or its parts in the case of a disconnected tree), one can convert all elements $\Omega$ into units using gauge transformations. Similarly, moving along each cycle, using gauge transformations, we can convert all elements $\Omega$ except one on each cycle to units. Thus the Statement 1 is proved.

Let $n_0$ and $n_1$ denote the numbers of vertices and 1-simplices of the graph, respectively.
The Appendix (\ref{A_Cycles_Number}) provides the following formula for the number of cycles of the graph:
\begin{gather}
C=n_1-n_0+1.
\label{Cycles_Number}
\end{gather}

Let us calculate the number of cycles of the graph of the complex $K^4_{pp'}$ (\ref{K^4pp'}). We have the number of vertices of this graph or complex:
\begin{gather}
n_0=r+2.
\label{K^4_n_0}
\end{gather}
Indeed, all vertices of the complex $K^4_{pp'}$ are exhausted by the set of vertices $\{a_{i\rho}\}$ plus two vertices $a_p$ and $a_{p'}$. The set of vertices $\{a_{i\rho}\}$ contains $r$ independent vertices.
The number of 1-simplices of the complex $K^4_{pp'}$ is equal to
\begin{gather}
n_{1K^4}=5(r-1).
\label{K^4_n_1}
\end{gather}
This number consists of $3(r-2)$ 1-simplices on $\partial K^3_p=S^2$ (\ref{Boundary}), plus $2r$ simplices $(a_{\alpha}a_p)$, $(a_{\alpha}a_{p'})$, $\alpha=1,\ldots,r$, plus the simplex $(a_pa_{p'})$. The number of 1-simplices on $\partial K^3_p$ is calculated as follows. Let $\tilde{n}_0=r$, $\tilde{n}_1$ and $\tilde{n}_2$ be the numbers of 0-,1- and 2-simplices on $S^2$, respectively. The Euler characteristic for $S^2$ gives the relation
$\chi=2=\tilde{n}_0-\tilde{n}_1+\tilde{n}_2=r-\tilde{n}_1+\tilde{n}_2$. On the other hand, on $S^2$ there is the relation $2\tilde{n}_1=3\tilde{n}_2$.
Eliminating $\tilde{n}_2$ from the last two equalities, we arrive at the equality of 1-simplices on $S^2$: $\tilde{n}_1=3(r-2)$. Using equalities (\ref{Cycles_Number}), (\ref{K^4_n_0}) and (\ref{K^4_n_1}) we find the number of cycles on the graph of complex $K^4_{pp'}$:
\begin{gather}
C_{K^4}=4r-6.
\label{Cycles_K^4}
\end{gather}
The number of 1-simplices of the complex $K^4_{pp'}$ on which the variables $\Omega=1$, according to (\ref{K^4_n_1}) and (\ref{Cycles_K^4}), is equal to
\begin{gather}
n_{\Omega=1}=n_{1K^4}-C_{K^4}=r+1.
\label{Omega=1_K^4}
\end{gather}
Taking advantage of the freedom to choose a detailed gauge fixation, we choose the 1-simplices $(a_{\alpha}a_p)$ and $(a_pa_{p'})$ as the (r+1) simplices on which $\Omega=1$.

Let us denote the WF on the 3D complex $\Sigma$ ($\Sigma'$)  as $\Phi_{\Sigma}$ ($\Phi_{\Sigma'}$).
As stated above, they are related by the following equation:
\begin{gather}
\Phi_{\Sigma'}=\int_{-\infty}^{-\infty}\prod_{\alpha=1}^r\d e_{a_{\alpha}a_p}
\d e_{a_Pa_{p'}}\int\d\overline{\Psi}_{a_p}\d\Psi_{a_p}
\nonumber \\
\times\exp(i\mA_{pp'})\Phi_{\Sigma}\equiv\exp(i\mA_{pp'})|\Phi_{\Sigma}\rangle.
\label{Evolution}
\end{gather}
The last identity is simply a designation.
Here the integral over the variables $\Omega$ is absent since on 1-simplices $(a_{\alpha}a_p)$ and $(a_pa_{p'})$ we have $\Omega=1$. In the Minkowski signature the integrand is oscillating, and therefore the integrals over tetrads in (\ref{Evolution}) converge. This is why the integration limits over tetrads are extended to infinity (compare with integral in (\ref{Partition_function})). \\

{\it Statement 2}. If the WF $\Phi_{\Sigma}$ is gauge invariant, then the WF $\Phi_{\Sigma'}$ is also gauge invariant. \\

Statement 2 is fairly clear: both the integrand and the measure of integration in (\ref{Evolution}) are gauge invariant. Therefore, the result of integration is also gauge invariant.

\section{Unitarity on a lattice}

\subsection{Unitarity of the local time evolution operator: First example}

Any WF defined on the complex $\Sigma$ is also defined on the complex $\Sigma'$, since these two complexes are identical. Therefore, we will not indicate further on which of these complexes the WF is defined.

Let us introduce some notations:
\begin{gather}
\langle\Phi_{\lambda_1}|\Phi_{\lambda_2}\rangle=\langle\Phi_{\lambda_2}|\Phi_{\lambda_1}\rangle^{\dag}\equiv
\nonumber \\
\int_{-\infty}^{+\infty}De\int D\mu\{\Omega\}_{\mbox{Gauge Fix}}\int D\overline{\Psi}D\Psi
\cdot\Phi^{\dag}_{\lambda_1}\Phi_{\lambda_2},
\label{Hilbert_Space}
\end{gather}
Here all integrals are taken over variables defined on the complex $\Sigma$ (or $\Sigma'$), and the gauge is fixed, which is indicated in the notation of the measure of variables $\Omega$. This operation is necessary since all WFs are gauge invariant.

We will need the following integral:
\begin{gather}
\mA_{pp'}|\Phi_{\lambda}\rangle
\nonumber \\
\equiv\int_{-\infty}^{-\infty}\prod_{\alpha=1}^r\d e_{a_{\alpha}a_p}
\d e_{a_Pa_{p'}}\int\d\overline{\Psi}_{a_p}\d\Psi_{a_p}\mA_{pp'}\Phi_{\lambda}.
\label{H_Local}
\end{gather}
It is useful to compare the right-hand sides of Eqs. (\ref{Evolution}) and (\ref{H_Local}).

We will assume that the set of WF $\{|\Phi_{\lambda}\rangle\}$ is an orthonormal basis (ONB):
\begin{gather}
\langle\Phi_{\lambda_1}|\Phi_{\lambda_2}\rangle=\delta_{\lambda_1\lambda_2}.
\label{ONB}
\end{gather}
Here the nature of the delta function on the right-hand side does not matter.

Further we consider only the normalizable WF $|\Phi\rangle=\sum_{\lambda}A_{\lambda}|\Phi_{\lambda}\rangle$:
\begin{gather}
\langle\Phi|\Phi\rangle=\sum_{\lambda}|A_{\lambda}|^2<\infty.
\label{Norma}
\end{gather}
Let us write out the second condition imposed on all normalizable WF:
\begin{gather}
|\langle\Phi_1|\mA_{pp'}|\Phi_2\rangle|<\infty.
\label{Limitation}
\end{gather}
We have:
\begin{gather}
\langle\Phi_1|\mA_{pp'}|\Phi_2\rangle^{\dag}=\langle\Phi_2|\mA_{p'p}|\Phi_1\rangle.
\label{Hermitianity}
\end{gather}
Here, as a result of conjugation, the complexes $\Sigma$ and $\Sigma'$ change places, which is taken into account in the designation of the operator $\mA_{pp'}^{\dag}=\mA_{p'p}$. However, using gauge transformations that do not change the integrals under consideration, the operator $\mA_{p'p}$ can be brought back to its previous form $\mA_{pp'}$, so that
\begin{gather}
\mA_{pp'}^{\dag}=\mA_{pp'}.
\label{Hermitianity_Fin}
\end{gather}

Since the operator $\mA_{pp'}$ is Hermitian on the considered space of functions, we will assume the ONB $\{|\Phi_{\lambda}\rangle\}$ to be a complete set of eigenfunctions of this operator:
\begin{gather}
\mA_{pp'}|\Phi_{\lambda}\rangle=E_{\lambda}|\Phi_{\lambda}\rangle, \quad |E_{\lambda}|<\infty, \quad
\overline{E}_{\lambda}=E_{\lambda}.
\label{EigenFunc}
\end{gather}

Let $|\Phi_1\rangle=\sum_{\lambda}A_{1\lambda}|\Phi_{\lambda}\rangle$ and
$|\Phi_2\rangle=\sum_{\lambda}A_{2\lambda}|\Phi_{\lambda}\rangle$. Then according to (\ref{ONB}) and (\ref{EigenFunc}) we have
\begin{gather}
 \langle\Phi_1|\mA_{pp'}|\Phi_2\rangle=\sum_{\lambda}\overline{A}_{1\lambda}A_{2\lambda}E_{\lambda},
\label{Hermit_ONB}
\end{gather}
\begin{gather}
\exp(i\mA_{pp'})\sum_{\lambda}A_{\lambda}|\Phi_{\lambda}\rangle=
\sum_{\lambda}A_{\lambda}e^{iE_{\lambda}}|\Phi_{\lambda}\rangle.
\label{Time_Evolution_5}
\end{gather}
From the last equality we see that as a result of the described time evolution (\ref{Evolution}) the coefficients $A_{\lambda}$ of the expansion of the WF in terms of the ONB $\{|\Phi_{\lambda}\rangle\}$ acquire phases: $A_{\lambda}\rightarrow A_{\lambda}e^{iE_{\lambda}}$. From this, in turn, it follows that the time evolution preserves both the norm (\ref{Norma}) of the vectors  and the values (\ref{Limitation}),
(\ref{Hermit_ONB}).

The above means the truth of the following Statement: \\

{\it Statement 3}. The operation of time evolution (\ref{Evolution}) is unitary and it preserves the class of functions on which unitarity is proven, as well as the number of degrees of freedom of the system.

\subsection{Further examples}

\subsubsection{Second example}

Let us consider the second example of local transformation of the WF in time. As in the first example, in this case a certain vertex $a_p\subset\Sigma$ transforms into a vertex $a_{p'}$ that does not belong to the complex $\Sigma$.
Here, the 4D complex transforming the complex $\Sigma$ into the complex $\Sigma'$ is denoted by
$K^4_{2pp'}$. The rest of the designations remain the same.  We believe that in this example $n=r=4$.
Four 3-simplices $(a_1a_2a_3a_p),(a_2a_3a_4a_p),(a_1a_3a_4a_p),(a_1a_2a_4a_p)\subset\Sigma$, and
$(a_1a_2a_3a_{p'}),(a_2a_3a_4a_{p'}),(a_1a_3a_4a_{p'}),(a_1a_2a_4a_{p'})\subset\Sigma'$.
By definition $K^4_{2pp'}=s^4_p\cup s^4_{p'}$, where 4-simplices $s^4_p=a_1a_2a_3a_4a_p$ and $s^4_{p'}=a_1a_2a_3a_4a_{p'}$.

The graph of the complex $K^4_{2pp'}$ contains $n_1=14$ 1-simplices and $n_0=6$ vertices. Therefore, according to Eq. (\ref{A_Cycles_Number}) this graph has 9 independent cycles, and there remain 5 1-simplices on which the variables $\Omega$ must be fixed. The corresponding connected tree subgraph containing all the original vertices of the graph may contain, for example, the following 1-simplices: $a_pa_1$, $a_pa_2$, $a_pa_3$, $a_3a_{p'}$, $a_{p'}a_4$.

The proof of the unitarity of the time evolution operator constructed here repeats verbatim the analogous proof given in the first example. The number of degrees of freedom is also conserved in the second example.

\subsubsection{Third example.}

Let us consider a 4D complex $K^4_{3pp'}$ with the following properties: $\partial K^4_{3pp'}=K^3_{3p}\cup K^3_{3p'}$, $K^3_{3p}\subset\Sigma$, $K^3_{3p'}\subset\Sigma'$, and the vertex $a_p$ ($a_{p'}$) is internal to the complex $K^3_{3p}$ ($K^3_{3p'}$). Moreover, the complex $K^4_{3pp'}$ is symmetric with respect to the inversion (reflection) operation $\hat{P}$ such that $\hat{P}K^3_{3p}=K^3_{3p'}$, $\hat{P}K^3_{3p'}=K^3_{3p}$, $\hat{P}a_p=a_{p'}$, $\hat{P}a_{p'}=a_p$, $\hat{P}K^4_{3pp'}=K^4_{3pp'}$,
$\hat{P}^2=1$. The $K^4_{pp'}$ and $K^4_{2pp'}$ complexes from the first two examples have these properties.

The evolution operator constructed on the complex $K^4_{3pp'}$ is also unitary and the number of degrees of freedom of the system is reproduced.

\subsubsection{Fourth example.}

Let, as above, $\Sigma$ and $\Sigma'$ be isomorphic 3D complexes. Let's consider a 4D complex $\mK$ with the following properties: (i) $\partial\mK=\Sigma\cup\Sigma'$ and (ii) in a finite or infinite number of steps on the complex $\mK$ such as those described in examples one, two and three, the complex $\Sigma$ is transformed into a complex $\Sigma'$. Since at each step the evolution of the wave function is unitary, and the evolution also conserves the number of degrees of freedom, then these same properties take place during the complete evolution from complex $\Sigma$ to complex $\Sigma'$. Of course, for this to be possible, the conditions (\ref{Limitation}) must be met at each vertex of the complex $\Sigma$. Note also that the result of the described evolution is determined uniquely up to a gauge transformation. The validity of the last statement follows from the fact that the value of the integral does not depend, generally speaking, on the order of integration. But the methods of fixing the gauge, although reflected in the final result, are not significant: the final WF is gauge invariant.

\section{Discussion}

{\bf 1.} Let us ask ourselves the first question that arises when studying the problem of unitarity: will the theory be unitary when the number of degrees of freedom changes in the process of evolution? Let us try to present some (sketchy and shaky) arguments in favor of the existing unitarity of such evolution. Here the argumentation will be fundamentally different from the one given above.

By definition, we will consider a linear operator ${\cal U}$, acting in a complex linear space with a scalar product, to be unitary if, as a result of its action on any vector, the scalar product of the resulting vector does not change:
\begin{gather}
|B\rangle={\cal U}|A\rangle, \quad \langle B|B\rangle=\langle A|A\rangle.
\label{Unitarity_Def}
\end{gather}
Here we admit that the vector $|B\rangle$ belongs to a space with a changed dimension. This is an expansive interpretation of the concept of unitarity, but it is precisely this interpretation that is needed here.

In our case, we denote the set of bosonic and fermionic variables on 3D complexes $\Sigma$ and $\Sigma'$ as $V$ and $V'$, respectively, and the transfer matrix transforming the WF $\Phi_A\{V\}$ on $\Sigma$ into WF $\Phi_B\{V'\}$ on $\Sigma'$, as $\exp i\mA\{V',V\}$. The transfer matrix is defined on one 4D lattice layer. Then
\begin{gather}
\Phi_B\{V'\}=\int_V\exp \big(i\mA\{V',V\}\big)\Phi_A\{V\},
\label{Evolution_II}
\end{gather}
and
\begin{gather}
\langle B|B\rangle=\int_{V_1}\int_{V_2}\int_{V'}\exp \Big(i[\mA\{V',V_1\}-\mA\{V',V_2\}]\Big)
\nonumber \\
\times \Phi^{\dag}_A\{V_2\}\Phi_A\{V_1\}.
\label{Evolution_Scalar}
\end{gather}
Since the lattice actions are Hermitian, the inner integral over the variables $V'$ is an alternating integral for $V_1\neq V_2$. The simplest model of an alternating integral is the integral
\begin{gather}
q=\left|\int_0^ae^{i\lambda x}\d x\right|=\left|\frac{2}{\lambda}\sin\frac{\lambda a}{2}\right|<a.
\label{Int_Alternating}
\end{gather}
Only for $\lambda=0$ we have $q/a=1$. From this we can conclude that the inner integral (\ref{Evolution_Scalar}) saturates at $V_1=V_2$. In other words, as a result of integration over variables $V'$, as the number of these integrations tends to infinity, the integral tends to the integral
\begin{gather}
\langle B|B\rangle=\int_{V_1}\int_{V_2}\Phi^{\dag}_A\{V_1\}\Phi_A\{V_1\}{\cal F}\{V_1\}\delta_{V_1V_2},
\nonumber \\
\delta_{V_1V_2}=0 \quad \mbox{for} \quad V_1\neq V_2, \quad \int_{V_2}\delta_{V_1V_2}=1.
\label{Transformation_3}
\end{gather}
Indeed, let us denote by $q_{\cW}$ the integral over the variables $V'$ of one 4-simplex $s^4_{\cW}$ at $V_1\neq V_2$. According to (\ref{Int_Alternating}) we have $q_{\cW}/(\mbox{measure volume})<1$. Therefore
\begin{gather}
\prod_{\cW}\left(\frac{q_{\cW}}{\mbox{measure volume}_{\cW}}\right)\longrightarrow 0
\label{Delta_Form}
\end{gather}
for an increasing number of integrations, that is, for an infinite transfer matrix in space.
If we could prove that the functional
\begin{gather}
{\cal F}\{V_1\}=\const,
\label{Unitarity_Condition}
\end{gather}
then the unitarity of the lattice theory would be proven for the case with a non-fixed number of degrees of freedom. Unfortunately, we do not know how to prove Eq. (\ref{Unitarity_Condition}) at present. Moreover, it is not clear to us whether this problem is correct.

The fermionic part of the actions in (\ref{Evolution_Scalar}) would not cause any significant difficulties in proving unitarity. The problem is in proving the equality (\ref{Unitarity_Condition}).

{\bf 2.} Let us briefly dwell on the problem of convergence of integrals when integrating over a non-compact subgroup of the gauge group $\Spin(3,1)$. We will denote the elements from this subgroup as ${\cal L}$:
\begin{gather}
{\cal L}=e^{\Vec{\sigma}\Vec{\chi}}=\cosh\chi+\sinh\chi(\Vec{\sigma}\Vec{n}),
\nonumber \\
\Vec{\chi}=\chi\Vec{n}, \quad  \Vec{n}=(\sin\theta\cos\phi,\,\sin\theta\sin\phi,\,\cos\phi),
\nonumber \\
0\leq\chi<\infty,
\label{Non-compact subgroup}
\end{gather}
and $\Vec{\sigma}$ are three Pauli matrices. The gauge invariant measure has the form
\begin{gather}
\d\mu({\cal L})=\frac{1}{12}\tr\Big\{\big({\cal L}^{-1}\d{\cal L}\big)\wedge{\cal L}^{-1}\d{\cal L}\big)\wedge
{\cal L}^{-1}\d{\cal L}\big)\Big\}
\nonumber \\
=\sinh^2\chi\sin\theta\d\chi\wedge\d\theta\wedge\d\phi.
\label{Measure}
\end{gather}
The oscillating exponential in the functional integral depends linearly on the value ${\cal L}$
in (\ref{Non-compact subgroup}). Integrating over the angles $\theta$ and $\phi$  eliminates the factor  $(\sinh\chi\sin\theta)$ from the measure (\ref{Measure}). Thus, the effective measure of integration over the variable $\chi$ becomes
\begin{gather}
\d\mu_{\mbox{eff}}=\sinh\chi\d\chi=\d\cosh\chi.
\label{Measure_Eff}
\end{gather}
Since the exponential contains a term of the form $(i\lambda\cosh\chi)$, the integral over variable $\chi$ turns out to be similar to the integral $\int\d x\exp(i\lambda x)$. The result of such integration leads to generalized functions that must be integrated during further integration.

{\bf 3.} The question arises about numerical modeling of the theory under study. It seems to us that the Monte Carlo method could be adequate here.

\section{Conclusion}

{\it The main conclusion of the work}: if space-time has the property of granularity, then it is possible to construct a quantum theory that models this granularity and has the property of unitarity.

The obtained result means that the lattice theory of gravity under consideration has the right to be considered a discrete regularization of the generally accepted continuous physical theory of gravity.

It should be noted that the lattice local transfer matrix constructed here has an analogue in the continuum theory of gravity. In the latter case, the Hamiltonian is an infinite linear combination of first-order constraints: ${\cal H}_T=\int c^{\Lambda}(x)\hat{\phi}_{\Lambda}(x)\d^3x$. Here $\hat{\phi}_{\Lambda}(x)$ are local first-class constraint operators, which are analogous to local transfer matrices in lattice theory. Physical states satisfy the conditions $\hat{\phi}_{\Lambda}(x)|\Phi\rangle=0$ (see
\cite{dirac2013lectures}, Lecture 3). The difference between the lattice and the continuum theory is that our consideration is carried out in the Schrödinger representation, while in the Lectures \cite{dirac2013lectures} --- in the Heisenberg representation.

In conclusion, we point out a modified model of discrete gravity, for which all the conclusions of this work remain valid. The modified model differs from the one considered only in that in the modified model the bosonic variables $e^a_{\cV_1\cV_2}$ are everywhere replaced by bilinear forms with respect to Dirac fields (see (\ref{Mink_5})):
\begin{gather}
e^a_{\cV_1\cV_2}\longrightarrow\Theta^a_{\cV_1\cV_2}.
\label{Replacement}
\end{gather}
This is possible because under the action of all symmetry transformations, both continuous and discrete, these quantities are transformed identically.

Note that the idea of representing a tetrad as a bilinear form of Dirac fields (quark-antiquark fields) has
long roots (see, for example, the works \cite{akama1978attempt,wetterich2004gravity,wetterich2022pregeometry,diakonov2011towards,vladimirov2012phase,
vladimirov2014diffeomorphism,obukhov2012extended}). In addition to the above-mentioned common property that unites the cited works, there are also significant differences between them. In particular, in the works \cite{akama1978attempt,wetterich2004gravity,wetterich2022pregeometry} continuum theories are studied and fundamental Lagrangians describing pregeometry are determined. The achievement of this approach is that it leads to Einstein's theory of gravity in the long-wave limit.
In the papers \cite{diakonov2011towards,vladimirov2012phase,
vladimirov2014diffeomorphism}, as in the present paper, the lattice regularization of the continuum theory of gravity is studied, and the lattice models coincide. In the paper \cite{obukhov2012extended} the ideas of the papers \cite{diakonov2011towards,vladimirov2012phase,vladimirov2014diffeomorphism} are developed, but within the framework of the continuum theory.

The replacement according to (\ref{Replacement}) simplifies the integration in the transition amplitude of type (\ref{Evolution}), since integrals over fermions, unlike integrals over bosons, always converge by definition. This approach was used in the works \cite{diakonov2011towards,vladimirov2014diffeomorphism}.
However, the problem of fixing the gauge in the Minkowski metric, as well as the problem of integration over the divergent volume of the gauge group $\Spin(3,1)$, still remain in the indicated modification of the theory.

\begin{acknowledgments}

We are grateful to one of our referees who insisted on the need to study the unitarity problem directly in lattice theory. We are grateful to A.Ya. Maltsev for significant assistance in solving mathematical issues, especially in the field of combinatorial topology.
This work was carried out as a part of the State Program FFWR-2024-0011.

\end{acknowledgments}

\appendix

\section{Some formulas from the basics of combinatorial topology}

Let us consider a one-dimensional abstract connected simplicial complex, which we will call a graph and denote by $\mK^1$. On each oriented 1-simplex $a_{{\cV}_1}a_{{\cV}_2}$ an element of the group $\Omega_{{\cV}_1{\cV}_2}=\Omega^{-1}_{{\cV}_2{\cV}_1}\in\Spin(3,1)$ is defined up to gauge transformations of the form (\ref{Gauge_Trans}) and $S_{\cV}\in\Spin(3,1)$. By definition, any two configurations of quantities $\{\Omega_{{\cV}_1{\cV}_2}\}_1$ and $\{\Omega_{{\cV}_1{\cV}_2}\}_2$ are equivalent if they are transformed into each other by gauge transformations alone. Next the set of equivalent configurations of quantities $\{\Omega_{{\cV}_1{\cV}_2}\}$ is called an equivalence class. Here the question of the number of equivalence classes on the graph is decided. The solution to this problem is necessary to fixing the gauge.

Let's call a closed contour
\begin{gather}
(a_{{\cV}_1}a_{{\cV}_2})(a_{{\cV}_2}a_{{\cV}_3})\ldots(a_{{\cV}_i}a_{{\cV}_1})
\label{Cycle}
\end{gather}
without self-intersections a cycle. Here 1-simplices $a_{{\cV}_i}a_{{\cV}_{i+1}}\in\mK^1$.

A graph is called tree if it does not contain cycles.

Let us show that on a tree graph the number of equivalence classes is equal to zero. Indeed, let us make the following gauge transformation: on the 1-simplex $a_{{\cV}_1}a_{{\cV}_2}$ we take $S_{{\cV}_1}=1$ and $S_{{\cV}_2}=\Omega_{{\cV}_1{\cV}_2}$. As a result of this gauge transformation, according to (\ref{Gauge_Trans}) we get $\tilde{\Omega}_{{\cV}_1{\cV}_2}=1$. In the next step on the 1-simplex $a_{{\cV}_2}a_{{\cV}_3}$, putting $S_{{\cV}_2}=1$ and $S_{{\cV}_3}=\Omega_{{\cV}_2{\cV}_3}$, we get $\tilde{\Omega}_{{\cV}_2{\cV}_3}=1$. Similarly, we can move in all directions from the initial 1-simplex $a_{{\cV}_1}a_{{\cV}_2}$. Since there are no cycles, there is no obstacle to achieving the result $\Omega=1$ on all 1-simplices. This means that on the tree graph there are no real degrees of freedom of the gauge field.

If the graph contains a cycle (\ref{Cycle}), then the described procedure is impossible.
Indeed, as a result of any gauge transformations, the value
\begin{gather}
\tr\Omega_{{\cV}_1{\cV}_2}\Omega_{{\cV}_2{\cV}_3}\ldots\Omega_{{\cV}_i{\cV}_1}
\label{Gauge_on_Cycle}
\end{gather}
is conserved. Thus, the problem of fixing the gauge is reduced to calculating the number of independent cycles of the graph. The definition of the set of independent cycles is given below. Although this set is not uniquely chosen, the number of independent cycles is a topological invariant.

Before we proceed to a theoretical consideration, let us demonstrate what has been said using a simple example. Consider a graph consisting of six tetrahedron edges $a_1a_2$, $a_2a_3$, $a_3a_1$, $a_1a_4$, $a_2a_4$, $a_3a_4$. Using the method described above, we can, for example, make $\Omega_{12}=\Omega_{23}=\Omega_{34}=1$. With the specified gauge fixing, three quantities
$\Omega_{13}$, $\Omega_{14}$, $\Omega_{24}$ remain independent variables. The stated choice of gauge fixing is ambiguous. In this example, the number of elementary cycles is four:
\begin{gather}
(a_1a_2)(a_2a_3)(a_3a_1), \quad (a_1a_2)(a_2a_4)(a_4a_1),
\nonumber \\
(a_2a_3)(a_3a_4)(a_4a_2), \quad (a_3a_1)(a_1a_4)(a_4a_3).
\label{Tetrahedron_Graph}
\end{gather}
But the number of independent cycles is three.

Let's move on to a general solution to the problem. For a more in-depth study of the problem, we recommend the reader the books \cite{pontryagin1976basics,dubrovin1984modern}.

In general, the number of independent degrees of freedom $\{\Omega\}$ is equal to the dimension (rank) of the one-dimensional homology group $H_1$ of the graph. By definition, this same number is the number of independent cycles on the graph. Let a base consisting of $C$ elements be selected in the group $H_1$.
Let us select one cycle $\mc$ in the graph representing each basis element in the group $H_1$.
Thus, the cycles $\mc_1,\ldots,\mc_C$ can be considered as a basis of the one-dimensional homology group of the graph. Generally speaking, the choice of base cycles is not straightforward. Each of these loops contains one independent variable $\Omega$, which is defined on some 1-simplex belonging to this loop.

Let's define a group $H_1$. Let $V_0$ denote the linear space of 0-dimensional chains. A 0-dimensional chain is a formal linear combination
\begin{gather}
\sum_{\cV}x^{\cV}a_{\cV}.
\label{0_chain}
\end{gather}
Here $x^{\cV}$ are real numbers and $a_{\cV}$ are graph vertices. Chains (\ref{0_chain}) can be multiplied by real numbers and added term by term. The linear space of 1-chains $V_1$ is defined similarly:
\begin{gather}
\sum_{1-simplices}x^{{\cV}_1{\cV}_2}(a_{{\cV}_1}a_{{\cV}_2}).
\label{1_chain}
\end{gather}
Here the coefficients of 1-simplices $a_{{\cV}_1}a_{{\cV}_2}$ are real numbers. The boundary operator on chains $\d:\,\,V_1\rightarrow V_0,\,\,V_0\rightarrow 0$ is linear and acts according to the rule
\begin{gather}
\d_1(a_{{\cV}_1}a_{{\cV}_2})=a_{{\cV}_2}-a_{{\cV}_1}.
\label{1_chain}
\end{gather}
The homology group $H_1$ is the kernel of the operator $\d_1$:
\begin{gather}
H_1=\mbox{Ker}\d_1, \quad \mbox{rank}H_1=\mbox{dim}\,\mbox{Ker}\d_1,
\label{H_1}
\end{gather}
"rank" means the dimension of linear spaces $H_1$ or $H_0$. We have for the Euler characteristic of the graph \cite{pontryagin1976basics,dubrovin1984modern}:
\begin{gather}
\chi=n_0-n_1=\mbox{rank}H_0-\mbox{rank}H_1.
\label{Euler_Char}
\end{gather}
Since by definition the number of independent cycles $C=\mbox{rank}H_1$ and for a connected graph we have $\mbox{rank}H_0=1$, then from here we get
\begin{gather}
C=n_1-n_0+1.
\label{A_Cycles_Number}
\end{gather}

To fix the gauge on the graph, you can do the following. We will sequentially throw out (mentally) one 1-simplex from each independent cycle. After each such act, the number of independent cycles of the original graph decreases by one. As a result of $C$ steps, the original graph becomes tree-like. On the remaining 1-simplices of the tree subgraph, we fix the variables $\Omega$, setting them equal to ones, or giving them other fixed values. On all mentally thrown out 1-simplices the variables $\Omega$ are independent and they take any values. Thus the gauge is fixed. The gauge fixation procedure is not unique.

Let's consider the first example from this point of view. Let us (mentally) throw out from the complex $K^4_{pp'}$ all 1-simplices, with the exception of the 1-simplices $(a_{\alpha}a_p)$ and $(a_pa_{p'})$.
The remaining graph becomes a a connected tree graph with the number of 1-simplices $n_1=(r+1)$ and the original number of vertices $n_0=(r+2)$. According to Eq. (\ref{A_Cycles_Number}) this graph has no cycles: $C=0$.

Let's go back to the example (\ref{Tetrahedron_Graph}). Since the tetrahedron graph has $n_1=6$ and $n_0=4$, then according to Eq. (\ref{A_Cycles_Number}) the number of independent cycles is three.


\end{document}